# Evidence for Suppression of Collective Magnetism in Fe-Ag Granular Multilayers

L.F. Kiss,* L. Bujdosó, and D. Kaptás

Wigner Research Centre for Physics, Budapest H-1525, P.O.Box 49, Hungary
*Corresponding author: L.F. Kiss
E-mail: kiss.laszlo.ferenc@wigner.hu

**Abstract**

Evidence for the suppression of collective magnetic behavior of dipolarly interacting Fe nanoparticles is found in Fe-Ag granular multilayers. Interaction of Fe particles separated by an Ag layer is studied as a function of the nominal thickness of the Ag layer. The surprisingly increasing interaction with increasing Ag-layer thickness, verified by memory-effect measurements, is explained by the formation of pinholes in the Ag layer at small Ag thicknesses, allowing direct ferromagnetic coupling between the Fe particles. This coupling may hinder the frustration of superspins favored by dipolar interactions. At larger Ag thicknesses, the Ag layer is continuous without pinholes and frustration leads to the appearance of the superspin-glass state. The effect of increasing interactions correlates well with the growing deviation at low temperatures of the measured field-cooled (FC) magnetization from the interaction-free FC curve calculated by a model based on the relaxation of two-level systems. Similar phenomenon is reported in a recently published paper (Sánchez et al., *Small* **2022**, 18, 2106762) where a dense nanoparticle system is studied. Here the collective magnetic behavior of the particles due to dipolar interactions is suppressed when the anisotropy energy of the individual particles exceeds a certain threshold.

## 1. Introduction

Fe-Ag multilayers of the sequence of Si/buffer + ($t_{Fe}$ Fe + $t_{Ag}$ Ag)$_n$ + cover have been shown to exhibit superparamagnetic (SPM) behavior of Fe islands within specific ranges of the $t_{Fe}$, $t_{Ag}$ and $n$ parameters [1]. Here $t_{Fe}$ and $t_{Ag}$ are the nominal thicknesses of the Fe and Ag layers, respectively, $n$ is the number of the Fe/Ag bilayers; “buffer” stands for a nonmagnetic metal layer evaporated directly on the smooth surface of the Si wafer (in our case it is Ag) and “cover” stands for a nonmagnetic layer which prevents the oxidation of the multilayer (here it is Nb). We studied [1] the magnetic properties of Fe-Ag multilayers fabricated with different $t_{\mathrm{Fe}}$, $t_{\mathrm{Ag}}$ and $n$ parameters each covering a wide range. We found that below a certain value of $t_{\mathrm{Fe}}$, $t_{\mathrm{Ag}}$ and $n$, the Fe layers are discontinuous and these granular multilayers show the magnetic behavior of a superparamagnetic ensemble. Besides that increasing $t_{\mathrm{Fe}}$ increases the blocking temperature, $t_{\mathrm{Ag}}$ and $n$ affect $T_{\mathrm{B}}$ not only by modifying the interactions between the Fe particles but also through influencing the growth process of Fe.

Intensive research into the interactions between magnetic nanoparticles has been done in the last two-three decades [2], [3].With increasing dipolar interactions, a transition sequence from superparamagnetic through super-spin glass (SSG) to superferromagnetic (SFM) behaviors could be observed [4], [5], [6], [7], [8], [9], [10], [11], [12], [13], [14], [15], [16], [17], [18], [19], [20]. Besides dipolar interactions, a Ruderman-Kittel-Kasuya-Yosida (RKKY) like exchange interaction between the nanoparticles was also proposed to be responsible for the collective SSG dynamics [21], [22], [23], [24], [25], [26]. In a recent paper [27], we investigated in detail the effect of interactions between the Fe particles on the SPM relaxation in Fe-Ag multilayers containing only one Fe layer with varying $t_{\mathrm{Fe}}$ and constant $t_{\mathrm{Ag}}$. Increasing strength of dipolar interactions was evidenced with increasing $t_{\mathrm{Fe}}$, probed by measuring the memory effect via the stop-and-wait protocol. We found an unambiguous correlation between the magnitude of the memory effect and the deviation of the low-field FC magnetization at low temperatures from the FC susceptibility curve of an interaction-free model of SPM particles.

Mørup proposed [28] a phase diagram of dipolarly interacting magnetic-particle systems where the collective freezing temperature is determined exclusively by the strength of the dipolar interactions. However, the role of the individual particle anisotropy on influencing this temperature remained an open question. Recently, Sánchez et al. studied [29] the influence of the individual particle anisotropy on the freezing temperature in a series of magnetic-particle assembly with similarly intense dipolar interactions but widely varying anisotropy. The anisotropy is tuned through different degrees of Co-doping in maghemite nanoparticles separated from each other by surfactants of different thicknesses. It was evidenced by experimental results and Monte Carlo simulations that the collective behavior (e.g., freezing temperature, memory effect, ac susceptibility) depends on both the dipolar interactions and the anisotropy-energy barrier of the nanoparticles. Moreover, they suggest that the collective magnetic behavior is suppressed below a crossover value of $T_{MAX}/T_B \approx 1.7$ (the ratio of the peak temperatures of the zero-field-cooled dc magnetization curves of dense and isolated systems made of the same nanoparticles) which corresponds to particle anisotropies above a threshold of $KV/E_{dd} \approx 130$ (the ratio of the anisotropy energy and dipolar interaction, $E_{dd}$, where $K$ is the anisotropy constant and $V$ is the particle volume).

In this paper, a Fe-Ag multilayer series was investigated containing two Fe layers with a nominal thickness of $t_{Fe}$ = 4 Å separated by an Ag layer of nominal thicknesses of $t_{Ag}$ = 8, 15, 26 and 50 Å ($n$ = 2) surrounded by buffer and cover layers. This series was fabricated in one evaporation run; therefore all layers of these samples except the varying Ag-layer separation were produced simultaneously under the same circumstances. The aim of this investigation is to explore the interactions between the Fe particles located in the two neighboring Fe layers. As a result, it will turn out that the $t_{Ag}$ dependence of the weak memory effect obtained cannot be interpreted solely by dipolar interactions but other factors (e.g., anisotropy) should be invoked, too, for the explanation.

## 2. Experimental Section

The multilayer samples were fabricated by vacuum evaporation. The detailed description of the equipment can be found elsewhere [1], [27]. We stress only that the two substrate holders and an appropriate shutter enable the preparation of two (in some cases four) samples with fully or partially identical layer structure.

The magnetization of the samples was measured using a MPMS-5S Quantum Design superconducting interference device (SQUID) between 5 and 300 K and 0 and 50 kOe. The low-field measurements were performed in $H$ = 10 Oe upon warming after cooling the samples from 300 to 5 K in zero field (ZFC) or in the measuring field of $H$ = 10 Oe (FC). The magnetization of the ZFC vs. temperature curves shown in this paper may assume negative magnetization values at low temperatures because of the remanent field (around –1 Oe) of the superconducting magnet (for detailed explanation see Refs.[1] and [27].

The measured ZFC and FC magnetization curves were compared with those of a theoretical model of superparamagnetic particle ensemble without interaction. The model is based on the fact that from energetic point of view an ensemble of single-domain particles can be described by an ensemble of two-level systems (TLS) [30]. The distribution of $V$ is approximated by a lognormal distribution

$$p(E_a) = \frac{1}{\sqrt{2\pi}\sigma E_a} \exp\left[-\frac{1}{2}\left(\frac{\ln\frac{E_a}{E_m}}{\sigma}\right)^2\right]$$

where $E_a = KV$ is the anisotropy energy, $E_m$ is the median, $\sigma$ is the width of the distribution and $K$ is assumed to be constant. The detailed description of the model can be found elsewhere [27].

In this paper, we use the stop-and-wait protocol as a genuine method to prove the existence of the SSG state of a nanoparticle system [8], [9], [10], [11], [13], [14], [19], [22], [26]. It means the measurement of the ZFC magnetization curve, stopping at a wait temperature ($T_w$) below the blocking (freezing) temperature ($T_B$), $T_w < T_B$, for a wait time ($t_w$) during cooling (called stop-and-wait curve). Subtracting from this curve the usual ZFC curve measured without stopping, we obtain the curve called memory effect. It shows a minimum close to $T_w$. The memory effect is called this way because the sample showing a dip at $T_w$ remembers the annealing at $T_w$ when it is reheated. Besides, a rejuvenation effect is also observed which means that after annealing at $T_w$ and subsequent cooling, when reheating, the low-field dc magnetization returns to the reference curve and deviates from it only close to $T_w$. Both features are impossible in an interaction-free SPM nanoparticle ensemble. In a simplified manner, the interactions can be considered as an internal field in which the particles are annealed. The wait temperatures, $T_w$, are selected to be around the inflexion point of the ZFC magnetization curve where considerable relaxation can be expected. The $T_w/T_B$ ratios for the presented memory-effect vs. temperature curves are similar (between 0.7-0.85) for the sake of quantitative comparability.

**2. Results**

Figure 1 shows the ZFC and FC magnetizations for the multilayer series where two Fe layers each with $t_{Fe}$ = 4 Å are separated by an Ag layer of varying thicknesses, $t_{Ag}$ = 8, 15, 26 and 50 Å (abbreviated by 2LFe-$t_{Ag}$ where $t_{Ag}$ = 8ÅAg, 15ÅAg, 26ÅAg and 50ÅAg). For comparison, the results for the 50 Å Ag + 4 Å Fe + 50 Å Ag + cover sample (with only one Fe layer, abbreviated by 1LFe-50ÅAg) are also plotted [27]. The solid lines are fits to the theoretical model of interaction-free superparamagnetic particle ensemble using lognormal distributions for the anisotropy energy ($E_a = KV$) with median ($E_m$) and width ($\sigma$) parameters given in Table I. The ZFC and FC magnetization vs. temperature curves shown in Fig. 1 follow the typical behavior of a SPM assembly. The qualitative description of this behavior is well known [1], [27]. The temperature of the ZFC peak is called the average blocking temperature of the sample ($T_B$). The SPM behavior is also evident from the magnetization curves as a function of the in-plane magnetic field measured above $T_B$, for which as two examples the *M*-*H* curves of the 2LFe-50ÅAg and 1LFe-50ÅAg multilayers are shown at some temperatures in Fig. 2. Because of the huge correction for the sample holder (diamagnetic Si wafer), the quality of these curves is poor.

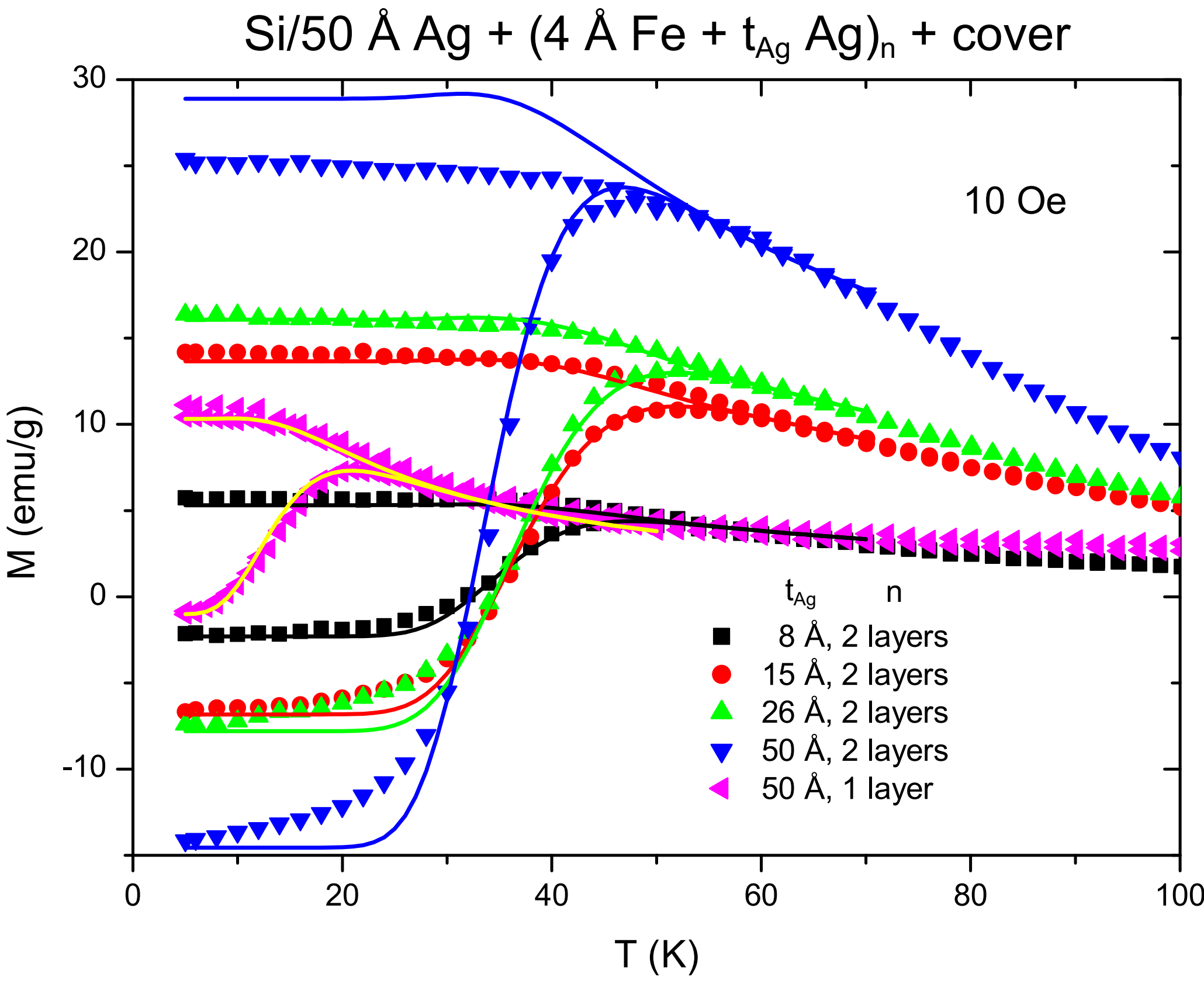

Fig. 1. ZFC (lower curves) and FC (upper curves) magnetization as a function of temperature for the (4 Å Fe +$t_{Ag}$ Ag)$_2$ + cover sample series ($t_{Ag}$ = 8, 15, 26 and 50 Å). For comparison, the curves for the 50 Å Ag + 4 Å Fe + 50 Å Ag + cover sample are also plotted [27]. The measuring field is $H$ = 10 Oe. The solid lines are fits to the theoretical model of interaction-free superparamagnetic particle ensemble (see Experimental section) using lognormal distributions for the anisotropy energy ($E_a = KV$) with median ($E_m$) and width ($\sigma$) parameters given in Table I. The small negative remanent field of the superconducting coil and the consequent negative magnetization values at low temperatures are taken into account with a slightly asymmetric initial population of the energy levels of the two-level systems (TLS), i.e., $n_0 > 0.5$ (see details in Experimental Section and Ref.[27]).

| $t_{Ag}$ (Å) | 8 | 15 | 26 | 50 |
|---|---|---|---|---|
| $E_m$ (eV) | 0.084 | 0.089 | 0.089 | 0.082 |
| $\sigma$ (eV) | 0.16 | 0.18 | 0.18 | 0.16 |
| $T_B$ (K) | 46 | 51 | 52 | 47 |
| $T_B/T_B^*$ | 2.19 | 2.43 | 2.48 | 2.24 |

Table 1 Anisotropy-energy median ($E_m$) and width ($\sigma$) of the lognormal distribution (see Experimental Section for definition) used for fitting the theoretical model of the interaction-free model of superparamagnetic particle ensemble to the ZFC and FC magnetization vs. temperature curves of the (4Å Fe +$t_{Ag}$ Ag)$_2$ + cover sample series shown in Fig. 1. $T_B$ is the blocking temperature (peak temperature of the ZFC curve). $T_B^*$ = 21 K is $T_B$ of the 50 Å Ag + 4 Å Fe + 50 Å Ag + cover sample regarded as the reference system without interaction [27].

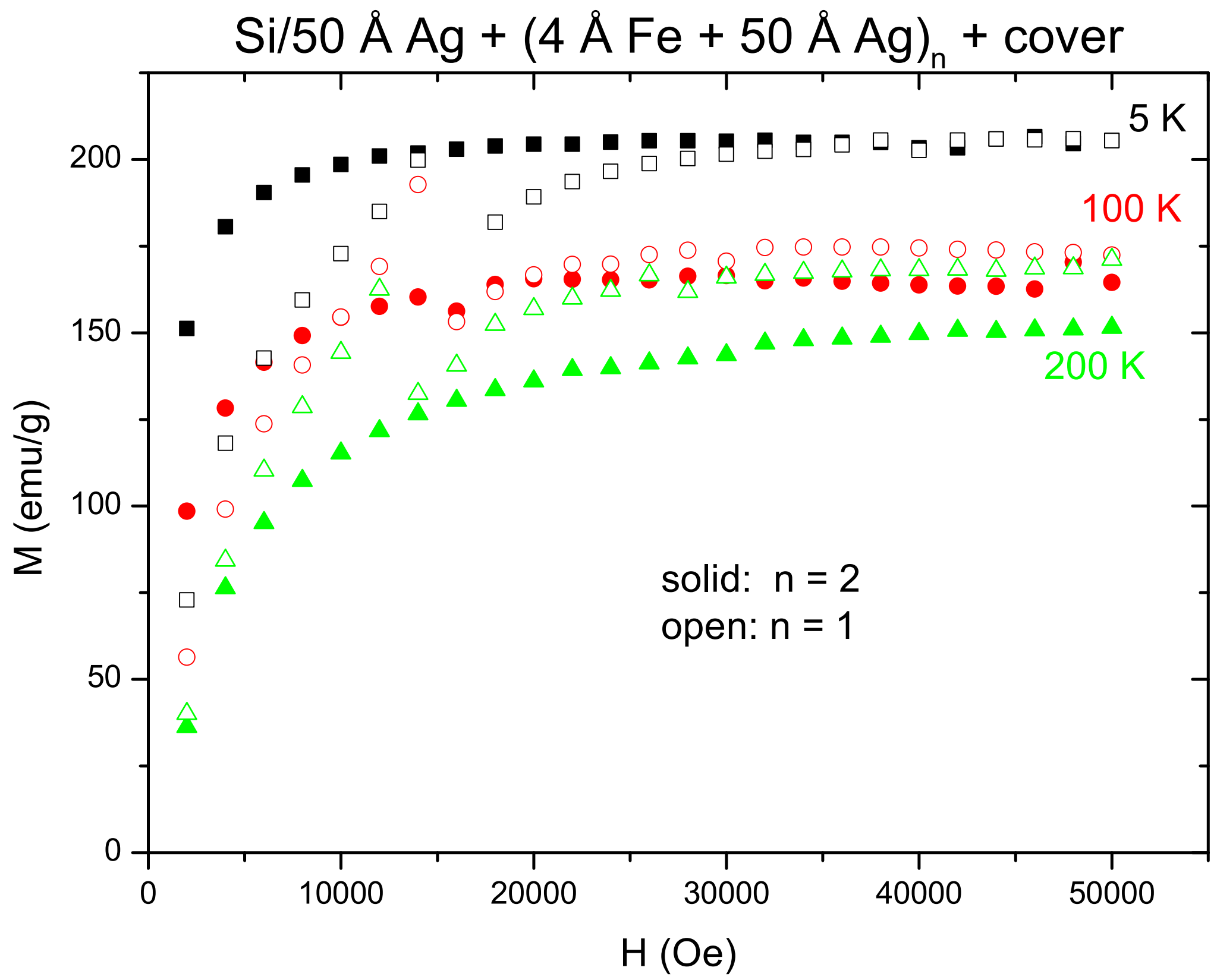

Fig. 2 Magnetization as a function of (in-plane) magnetic field for the (4 Å Fe + 50 Å Ag)$_n$ + cover multilayers with $n$ = 1 (open symbols) and $n$ = 2 (solid symbols) at $T$ = 5, 100 and 200 K as indicated by labels.

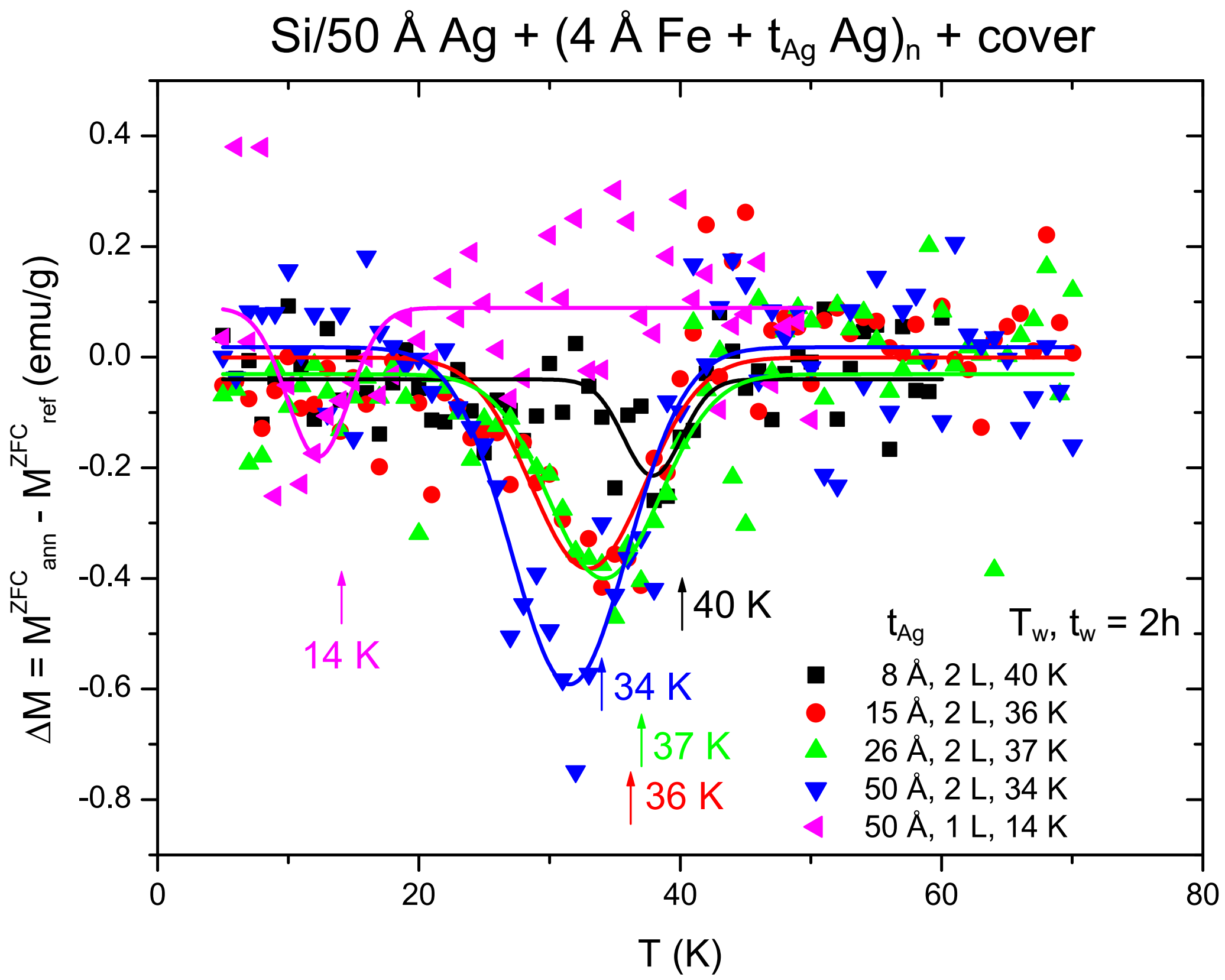

Fig. 3 Memory effect for the (4Å Fe +$t_{Ag}$ Ag)$_2$ + cover sample series ($t_{Ag}$ = 8, 15, 26 and 50 Å) with wait time, $t_w$ = 2h, and wait temperatures, $T_w$, indicated in the figure. For comparison, the curve for the 50 Å Ag + 4 Å Fe + 50 Å Ag + cover sample is also plotted [27]. The ZFC magnetization both in the annealed and the reference states as a function of temperature was measured in $H$ = 10 Oe. The arrows and temperatures denoted by labels show the wait temperatures, $T_w$, where the respective sample was annealed for $t_w$ = 2h. L means layer and the lines are Gauss fits used as guides to the eye.

As seen in Fig. 1, the blocking temperature is doubled when the bilayer number is increased from $n$ = 1 to 2 for the sample with $t_{Fe}$ = 4 Å and $t_{Ag}$ = 50 Å. As it will be evident below, this increase in $T_B$ is mainly attributed to the dipolar interactions between the Fe particles and only partially to the increase in cluster size. It is because the cluster size is dominantly determined by the nominal Fe thickness ($t_{Fe}$) [27] and it was kept constant ($t_{Fe}$ = 4 Å) in the 2LFe-$t_{Ag}$ multilayer series studied here. Some increase in cluster size can be explained [1] by the changing morphology during the columnar-structure growth of the layers preferring larger cluster sizes with increasing $n$. However, decreasing the Ag-layer thickness from $t_{Ag}$ = 50 to 8 Å, $T_B$ remains practically constant. This result is surprising for two reasons: (i) we obtained previously [1] that $T_B$ increases with decreasing $t_{Ag}$ for a multilayer series with $t_{Fe}$ = 2 Å and $n$ = 75; and (ii) the dipolar interactions between Fe clusters being in different Fe layers are expected to increase with decreasing $t_{Ag}$, resulting in a higher $T_B$. The measured $T_B$ values can be regarded reliable since this sample series was fabricated in one evaporation cycle.

Figure 3 shows the memory effect for the same samples as in Fig. 1 with wait time, $t_w$ = 2h, and wait temperatures, $T_w$, indicated in the figure. The memory effect is almost absent for the $t_{Ag}$ = 8 Å sample with two Fe layers ($t_{Fe}$ = 4 Å, $n$ = 1) and also for the 1LFe-50ÅAg sample containing a single Fe layer. Though a tiny dip close to $T_w$ can be surmised for both samples as shown by the Gauss fits (used as guides to the eye), the presence of memory effect cannot be unambiguously stated because of the large scattering of the measuring points. The memory effect is very small but clearly observable for $t_{Ag}$= 15 and 26 Å and it is definitely larger for $t_{Ag}$ = 50 Å of the sample series consisting of two Fe layers.

Similarly as found for the 50 Å Ag + $t_{Fe}$ Fe + 50 Å Ag + cover sample series with a single Fe layer having nominal thicknesses of $t_{Fe}$ = 4, 5, 7 and 10 Å [27], a correlation is observed in Figs. 1 and 2 between the behavior of the FC magnetization and the memory effect for the presently studied 2LFe-$t_{Ag}$ sample series. In these two-Fe-layer sample series, there is a systematic deviation of the theoretical FC curve from the measured one below $T_B$ for larger $t_{Ag}$ while the theoretical ZFC curves describe relatively well the measured ones for all $t_{Ag}$. For larger $t_{Ag}$, the measured FC magnetization remains below the calculated curve (for $t_{Ag}$ = 50 Å significantly), becoming more and more flattened without exhibiting any plateau at low temperatures. The correlation found between the appearance of the memory effect and the flattening of the FC magnetization curve was previously attributed to the appearance of dipolar interactions between the Fe clusters [27]. Similar observations were reported by Binns et al. [31]. and Denardin et al. [32]. The rising of the theoretical FC curve at low temperatures is a consequence of the non-collective behavior of the clusters, causing the freezing of the individual magnetic moments upon cooling, independently from each other, aligned to the direction of the field. The comparison of the measured and theoretical FC magnetization curves leads to the conclusion that the multilayer sample with $t_{Ag}$ = 8 Å can be well described by the interaction-free model. Increasing $t_{Ag}$, the interactions should increase as is obvious from the increasing deviation between the measured and model FC curves at low temperatures (Fig. 1). This surprising trend is corroborated by the results of the memory-effect measurements (Fig. 3) proving that the interlayer interactions should increase with increasing $t_{Ag}$.

It is obvious that the increasing memory effect with increasing $t_{Ag}$ cannot be explained exclusively by the dipolar interactions between the Fe clusters. There must be another effect in play which counteracts the tendency of the dipolar interactions to frustrate the magnetic moments at small $t_{Ag}$.

## 3. Discussion

For the multilayer series with two nominal Fe layers, 2LFe-$t_{Ag}$, a surprising result is obtained for the $t_{Ag}$ dependence of the memory effect (Fig. 3), supported also by the behavior of the low-field FC magnetization at low temperatures (Fig. 1): the memory effect, i.e., the interactions between the clusters increase with increasing Ag-layer thickness. Since the presence of dipolar interactions only would cause an opposite dependence, we have to assume other factors, too, influencing the possibility of the superspins to become frustrated, i.e., to form an SSG state.

Besides the dipolar interaction, it is the RKKY exchange interaction which might cause an SSG state of the superspins in granular alloys in which the magnetic particles are separated by a non-magnetic metallic (conductive) matrix [21], [22]. [23], [24], [25]. In these cases, a small part of Co is dissolved in Ag [21], [22] and Cu [23] matrix or a small part of Fe in Cu matrix [25] where the dissolved magnetic atoms or groups containing a few of them strengthen the effect of the RKKY interaction between the magnetic particles. It is obvious that this mechanism could not work in Fe/Ag systems since the solubility of Fe in Ag is very small compared to the previous cases. Intermixing of Fe and Ag could be verified only in two-monolayer thickness at the interface between Fe and Ag layers of a Ag/Fe/Ag trilayer with the help of magnetoresistance measurements [33]. Moreover, the RKKY interaction would lead to an opposite dependence of the memory effect with increasing Ag-layer thickness as was observed for our 2LFe-$t_{Ag}$ multilayer series (Fig. 1). Since the RKKY interaction significantly weakens with increasing $t_{Ag}$ in the studied $t_{Ag}$ range, the memory effect should decrease with increasing $t_{Ag}$, were it caused by this interaction.

Despite all of this, an oscillatory RKKY indirect exchange coupling of ferromagnetic layers (e.g., Co) via non-magnetic spacer layers (e.g., Cu) exists and is verified in multilayers even with contiguous (not granular) layer structure [34]. This oscillating coupling, i.e., alternating ferromagnetic (FM) and antiferromagnetic (AFM) interlayer interaction as a function of the thickness of the non-magnetic spacer layer, can be demonstrated by magnetoresistance measurements but magnetization studies often fail to prove the existence of AFM couplings [35], [36]. It was shown [35], [36], [37] that structural imperfections (e.g., pinholes) in the spacer layers give rise to direct FM coupling of neighboring ferromagnetic layers.

The discontinuity of an Ag layer below a thickness of 50 Å was also shown by Mössbauer spectroscopy in B/Fe/Ag multilayer samples [38]. The blocking temperature of one nominally 4-Å thick Fe granular layer evaporated on an Ag layer of varying thicknesses shows a minimum at $t_{Ag}$ = 40-70 Å which was also explained by the onset of such discontinuities called pinholes below a critical thickness at $t_{Ag}$ ~50 Å [1]. This observation offers a plausible interpretation for the anomalous behavior of the Ag-thickness dependence of the memory effect in the 2LFe-$t_{Ag}$ multilayer series (Fig. 3). At small $t_{Ag}$, the Fe particles located in the neighboring Fe layers become FM coupled via pinholes in the Ag layer separating them, hindering the frustration of the superspins, hence the appearance of the SSG state. With increasing $t_{Ag}$, this effect gradually ceases as the Ag layer becomes contiguous, and the dipolar interactions between the Fe clusters (or at least between parts of them) are able to form an SSG state evidenced by the memory effect measured. The RKKY interaction would not be effective at Ag thicknesses close to $t_{Ag}$ = 50 Å. The observation that the blocking temperature does not vary with $t_{Ag}$ in this multilayer series (Fig. 1) might be the consequence of a delicate balance of both effects. At small $t_{Ag}$, $T_B$ is increased with respect to the one-Fe-layer case ($n$ = 1) because of the direct contact of Fe particles via pinholes resulting in larger average particle volume through coalescence while at larger $t_{Ag}$ in absence of this effect, the $T_B$ increase is caused by the dipolar interactions, apparently increasing the average particle volume. The previously obtained increasing $T_B$ with decreasing $t_{Ag}$ in the same $t_{Ag}$ range for the $t_{Fe}$ = 2 Å and $n$ = 75 multilayer series [1] apparently contradicts the above observed finding. However, we found evidence in the same paper for the increase of the Fe-particle size along the multilayer stack as the subsequent layers are evaporated. The significant $T_B$ increase with increasing bilayer number measured in Fe-Ag multilayers [1] was partly explained by this particle-size increase. We think this is the reason for the different behavior of the $T_B$ vs. $t_{Ag}$ dependence when the bilayer number is increased from $n = 2$ to 75.

For the (26 Å Ag + $t_{Fe}$ Fe)$_{10}$ multilayer series (2 Å $\leq t_{Fe} \leq$ 10 Å), a change in the direction of the spontaneous magnetization from out-of-plane to in-plane was observed with increasing $t_{Fe}$ at $t_{Fe}$ ~6 Å by Mössbauer spectroscopy [39]. Since the appearance of perpendicular anisotropy below $t_{Fe}$ ~6 Å seems to be an intrinsic property of Fe-Ag multilayers, an out-of-plane anisotropy can be assumed also for the 2LFe-$t_{Ag}$ multilayer series studied here. We think that at small $t_{Ag}$, the perpendicular anisotropy might strengthen the effect of pinholes in hindering the frustration of the superspins.

The mechanism which we used to explain the suppression of collective magnetic behavior for decreasing thickness of the Ag-layer surrounded by two Fe layers in the studied Fe-Ag granular multilayer series is very similar to the interpretation of Sánchez et al. [29]. In their nanoparticle system, the particle volumes are almost constant (with narrow particle-size distribution) and the anisotropy constant is increased by nearly an order of magnitude with Co-doping. In our case, the anisotropy constant does not vary significantly but the particle volumes should increase by contacts through pinholes (coalescence) for decreasing $t_{Ag}$. In both cases, the anisotropy energy ($KV$) increases, leading to the suppression of collective magnetism. In our case, the 1LFe-50ÅAg sample containing only one Fe layer can be approximately regarded as the reference sample without interaction. This sample cannot be exactly considered as a true reference since the cluster sizes for the 2LFe-$t_{Ag}$ sample series may be somewhat larger than for the 1LFe-50ÅAg reference sample while the dilute (i.e. reference) and concentrated samples of Sánchez et al. [29] consist of the same particles. The ratio of the peak temperatures of the ZFC magnetization curves of the two-Fe-layer samples and the one-Fe-layer reference sample, $T_B/T_B^*$ (which corresponds to $T_{MAX}/T_B$ in Ref.[29]) varies between 2.19 and 2.48 (Table 1). These values are somewhat larger than the threshold of 1.7 found by Sánchez et al. [29]. The reason for this difference might be that the nanoparticle system in the Fe-Ag granular multilayers is not as ideal as that studied in Ref.[29]. As it was pointed out in a previous paper [1], the increase of $T_B$ with increasing bilayer number in Fe-Ag granular multilayers is partly related to the increase of the particle size in successive Fe layers during the evaporation process. Therefore, the particle volumes are not monodisperse; instead, the particle size can almost be doubled in the second Fe layer with respect to the first one. Broader particle-size distributions require higher ratios of $T_\mathrm{B}/T_\mathrm{B}^*$ to reach collective behavior [29].

The measurement of the ZFC and FC magnetization curves of a dense nanoparticle system, fitted with the help of the proposed model of an interaction-free SPM particle assembly is an easy experimental method to assess the collective character of magnetic behavior of the system.

## 4. Conclusions

Superparamagnetic Fe-Ag granular multilayers where the nominal thickness of a Ag layer located between two nominally 4-Å thick Fe layers is increased from $t_\mathrm{Ag}$ = 8 to 50 Å were investigated by measuring the ZFC and FC magnetization curves and the memory effect with the help of the stop-and-wait protocol. For increasing $t_\mathrm{Ag}$, an increase of the memory effect was observed despite the decreasing dipolar interactions. We found an unambiguous correlation between the magnitude of the memory effect and the deviation of the low-field FC magnetization at low temperatures from the FC susceptibility curve of an interaction-free model of SPM particles. The suppression of collective magnetic behavior for decreasing $t_\mathrm{Ag}$ is explained by the formation of pinholes in the Ag layer at small thicknesses, bringing about ferromagnetic coupling between the Fe particles in adjacent Fe layers, which may impede the frustration of the superspins.

## Acknowledgements

The Wigner Research Centre for Physics utilizes the research infrastructure of the Hungarian Academy of Sciences and is operated by the Eötvös Loránd Research Network (ELKH) Secretariat (Hungary). Financial support by the Hungarian Scientific Research Fund (Grant No. OTKA-K-112811) is greatly acknowledged. The authors are grateful to Dr. I. Bakonyi for valuable discussions.